\documentclass[twocolumn,floatfix]{revtex4-1}
\pdfoutput=1
\usepackage{color} 
\usepackage{graphicx} 
\usepackage{amsmath,amssymb}

\newcommand{\red}[1]{\textcolor{black}{#1}}

\begin{document}

\author{Roman Shevchuk } 
\author{Diego Prada-Gracia} 
\affiliation{ Freiburg Institute for Advanced Studies, University of Freiburg, Freiburg, Germany.}

\author{Francesco Rao \footnote{francesco.rao@frias.uni-freiburg.de}}
\affiliation{ Freiburg Institute for Advanced Studies, University of Freiburg, Freiburg, Germany.}

\title{Water structure-forming capabilities are temperature shifted for different models}

\begin{abstract} 

A large number of water models exists for molecular simulations. They differ in
the ability to reproduce specific features of real water instead of others,
like the correct temperature for the density maximum or the diffusion
coefficient.  Past analysis mostly concentrated on ensemble quantities, while
few data was reported on the different microscopic behavior.  Here, we compare
seven widely used classical water models (SPC, SPC/E, TIP3P, TIP4P, TIP4P-Ew,
TIP4P/2005 and TIP5P) in terms of their local structure-forming capabilities
through hydrogen bonds for temperatures ranging from 210 K to 350 K by the
introduction of a set of order parameters taking into account the configuration
of the second solvation shell.  We found that all models share the same
structural pattern up to a temperature shift.  When this shift is applied, all
models overlap onto a master curve.  Interestingly, increased
stabilization of fully coordinated structures extending to at least two
solvation shells is found for models that are able to reproduce the correct
position of the density maximum. Our results provide a self-consistent
atomic-level structural comparison protocol, which can be of help in
elucidating the influence of different water models on protein structure and
dynamics.

\end{abstract}

\date{\today}

\maketitle

\section{Introduction}

At the fundamental level, water directly influences several
biologically relevant processes including protein folding
\cite{Levy2006}, protein-protein association
\cite{Papoian2003,Levy2006,Hummer2010,Ahmad2011} and amyloid
aggregation \cite{Thirumalai2011}. Surprisingly, relatively simple models with fixed charges and geometry are able to reproduce the phase diagram as well as many of the anomalies of water with good accuracy \cite{Abascal2009,Aragones2011}. For example, all popular classical water models present a density maximum \cite{Vega2005,Abascal2005}. However, only those that explicitly included this information in the fitting of the potential are able to correctly reproduce the experimental value located at around 277~K at ambient pressure \cite{Kell1975}.  

Due to their improved speed, biomolecular simulations in explicit water were traditionally run with TIP3P\cite{Jorgensen1983} or SPC\cite{Berendsen1981}. Nowadays, more elaborated models can be easily used and their impact on the calculation assessed \cite{Florova2010}. Optimized four site models reproducing the experimental temperature of maximum
density seem to improve the quality of biomolecular simulations. Best and collaborators showed that predicted helical propensities are in
better agreement with experiments when a TIP4P/2005 water model is
chosen in place of the traditional TIP3P \cite{Best2010}. Others
reported that TIP4P-Ew provides better free-energy
estimations compared to conventional water models \cite{Nerenberg2011}. In both studies, the improved behavior was not connected to a clear microscopic property of the water model. To this aim, one limitation is the lack of a common framework to compare the structural behavior of liquid water at the atomic level.

Here, seven popular classical water models, namely SPC \cite{Berendsen1981},
SPC/E \cite{Berendsen1987}, TIP3P \cite{Jorgensen1983}, TIP4P
\cite{Jorgensen1985}, TIP4P-Ew \cite{Horn2004}, TIP4P/2005 \cite{Abascal2005}
and TIP5P \cite{Mahoney2000} are investigated in terms of their local structure
forming capabilities. That is, their ability to form structured or partially
structured environments of the size of up to two solvation shells through
hydrogen bonds.  This approach represents a simplified version of the recent
complex networks analysis introduced to study water structural inhomogeneities
\cite{Rao2010,Garrett2011}.

\begin{figure}[t]
  \includegraphics[width=40mm]{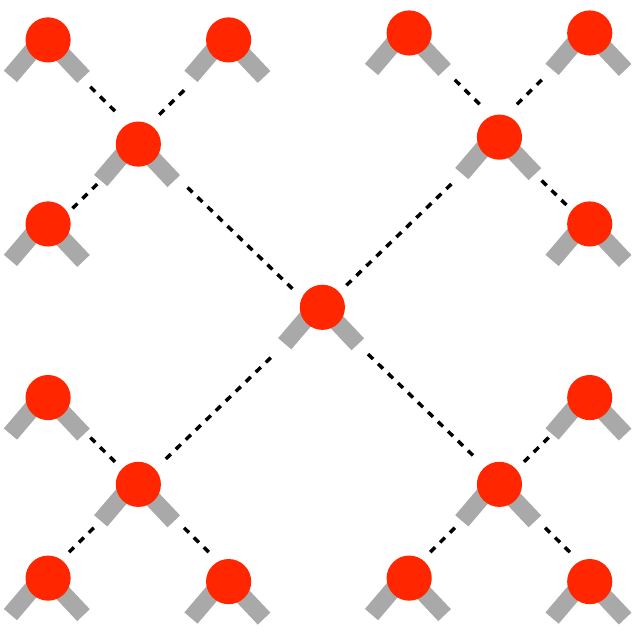}
  \caption{Schematic representation of the fully coordinated water configuration (P$_4$ population, see text). Dashed lines represent hydrogen bonds}
  \label{fig:p4}
\end{figure}

\section{Methods}

\subsection{Simulations}

Molecular dynamics simulations were run with the program GROMACS
\cite{gromacs} with an integration time-step of 2~fs. The water box
consisted of 1024 water molecules in the NPT ensemble with pressure of
1 atm and temperatures ranging from 210 K to 350 K with steps of 10 K.
The Berendsen barostat \cite{Berendsen1984}, velocity rescale
thermostat \cite{Bussi2007} and PME \cite{Darden1993} were used for
pressure coupling, temperature coupling and long-range electrostatics,
respectively.

The present analysis was done over 25'000 snapshots obtained
from a 100~ps long run after a 10~ns equilibration in the same
conditions. The short length of the analyzed trajectory is justified
because structural properties are calculated for each water
molecule, effectively improving the sampling by three order of
magnitudes (there are 1024 water molecules). Repeating the analysis
with a longer equilibration time reproduced the same results. TIP5P
data was collected up to 230~K, just before the approaching of the
glass-transition \cite{Brovchenko2005}.

\subsection{Density maximum}

The position of the maximum density was obtained from 1~ns long
simulations after 10~ns of equilibration in the NPT ensemble.  The
temperature of maximum density was extracted by polynomial fitting
around the maximum.  Variations from the literature (see
Table~\ref{table:density}) may be due to size effects and a different
treatment of the electrostatics. The location of the TIP3P density
maximum was obtained by running further simulations at lower
temperatures.

\subsection{Energetics and tetrahedral order parameter}

The free energy of a configuration $i$ is given by 

\begin{equation}
\Delta F_{i}=-k_B T \log(P_{i}),
\end{equation}

where $k_B$ is the Boltzmann factor, $T$ the temperature and $P_{i}$ the
population of the selected configuration. The enthalpy is estimated by summing
up all pairwise contributions between the water molecules belonging to the same
configuration.

The tetrahedral order parameter for a water molecule $i$ is calculated as

\begin{equation}
q_{i}=1-\frac{3}{8}\sum_{j=1}^{3}\sum_{k=j+1}^{4}\big(cos\psi_{jik}+\frac{1}{3}\big)^{2},
\end{equation}
where $j$ and $k$ are any of the four nearest water molecules of $i$
and $\psi_{jik}$ is the angle formed by their oxygens\cite{
  Errington2001}. The averaged value of this order parameter over an
ensemble of water molecules is denoted as $Q$.

\section{Results}

\begin{figure*}[H]
  \includegraphics[width=160mm]{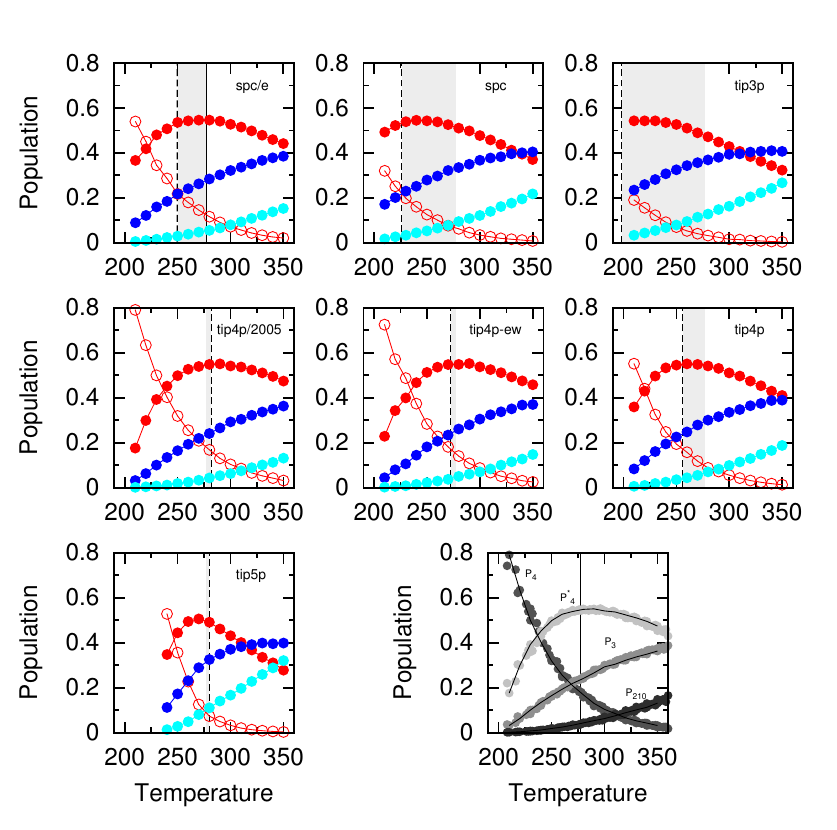}
  \caption{Temperature dependence of water structure populations for seven
  classical water models.  P$_4$, P$_4^*$, P$_3$, and P$_{210}$ are shown in
  red empty, filled red,  blue, and light blue circles, respectively
  (see text for details).  The gray stretch highlights the temperature
  difference between the calculated position of the temperature of maximum
  density (vertical dashed line, see also Table~\ref{table:density}) and the
  experimental value at 277~K (solid line).  The bottom right monochrome plot
  shows the superposition of all models after temperature shifting each data
  set (TIP4P/2005 data was used as reference). For each
  temperature, the sum over the 4 groups is equal to one.  }
  \label{fig:micro_pop}
\end{figure*}

\subsection{Structure forming capabilities}

Water structure forming capabilities were investigated by analyzing
the hydrogen-bond network of each water molecule in the simulation box
together with its first and second solvation shells. A maximum of four
hydrogen-bonds per molecule was considered. A bond is
formed when the distance between oxygens and the angle O-H-O is
smaller than 3.5 \AA\ and 30 degrees, respectively~\cite{Luzar1996}. Water
structures were grouped into four archetypal configurations of
population P$_i^{(*)}$: the fully coordinated first and second
solvation shells for a total of 16 hydrogen-bonds (P$_4$, see
Fig.~\ref{fig:p4} for a schematic representation); the fully
coordinated first shell, in which one or more hydrogen bonds between
the first and the second shells are missing or loops are formed
(P$_4^*$); the three coordinated water molecule (P$_3$) and the rest
(P$_{210}$). Within this representation the sum over the four
populations is equal to one for each temperature.

In Fig.~\ref{fig:micro_pop}, the temperature dependence of the four
microscopic water structures is shown.  Among the different water
models, the qualitative behavior is strikingly similar.  Three main
types of temperature scalings were observed: increasing population
with decreasing temperature (enthalpically stabilized); increasing
population with increasing temperature (entropically stabilized); with
a maximum, where a turnover between enthalpic and entropic
stabilization takes place at a model dependent temperature. All four
water configurations fall into one of these three main classes.  The
population of the fully ordered structure, P$_4$, increases with
decreasing temperature (Fig.~\ref{fig:micro_pop}, red empty
circles). Consequently, this configuration is enthalpically
stabilized.  This is not the case when defects in the hydrogen bond
structure are introduced (P$_4^{*}$, filled red circles). For this
configuration the population increases with decreasing temperature
until it reaches a maximum in correspondence to a rapid increase of
the population of the fully-coordinated configuration. The maximum is
located in a temperature range close to the temperature of maximum
density of the model under consideration (dashed vertical line).
Finally, both P$_3$ and P$_{210}$ are mainly entropically stabilized,
showing larger populations at higher temperatures. Taken together,
these results indicate that specific water configurations dominate at
each temperature range: full-coordination extending to at least two
solvation shells at low temperatures, four-coordinated configurations
with no spatial extension at intermediate temperatures and mainly
disordered ones at higher temperatures.

\begin{figure}[t]
  \includegraphics[width=80mm]{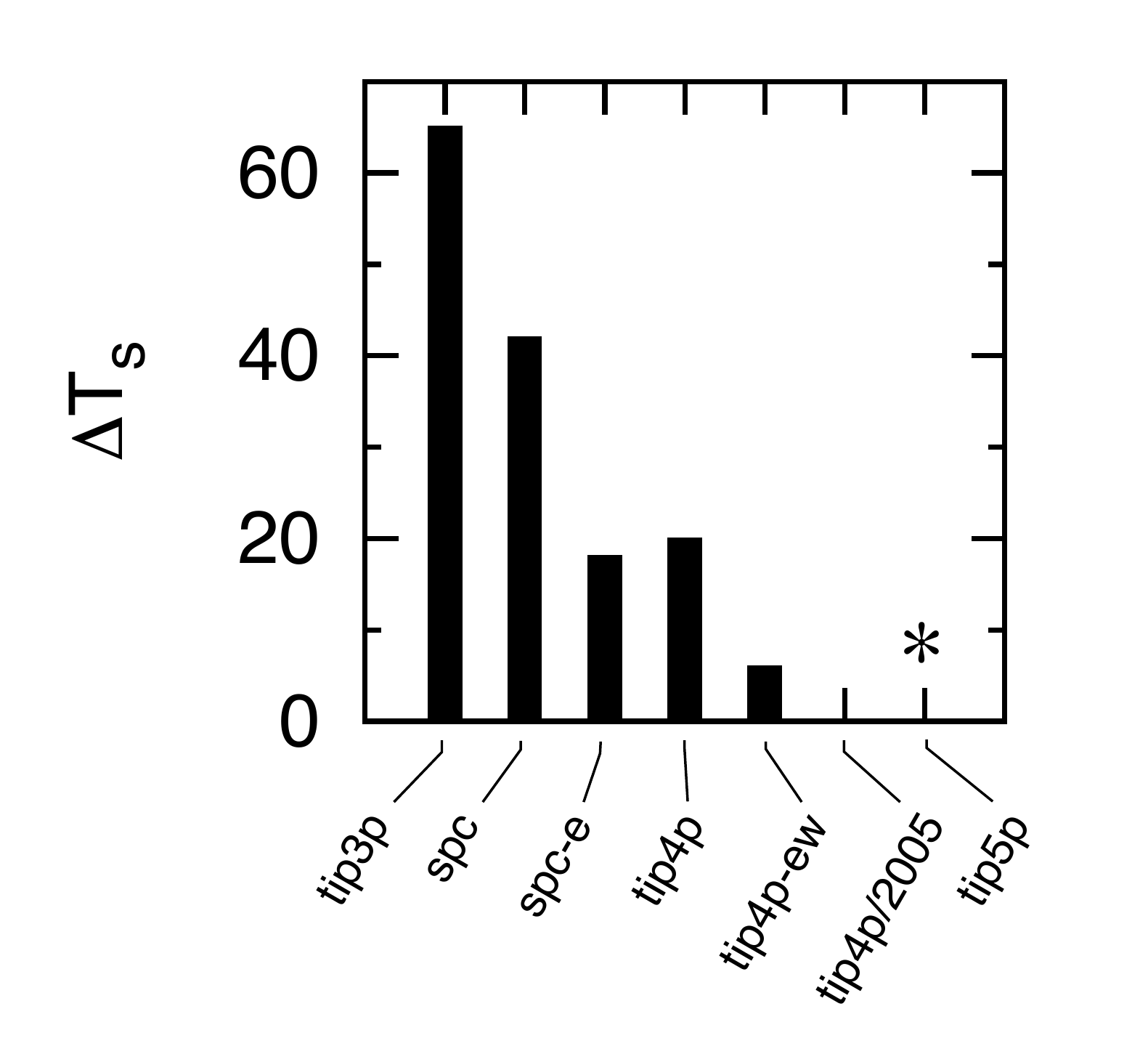}
  \caption{Structural temperature shift $\Delta T_s$ with respect to the
  TIP4P/2005 model. TIP5P was excluded from the superposition analysis (see
  text for details) .}
  \label{fig:deltats}
\end{figure}

Despite these similarities, an important difference among the models is the
temperature range at which the relative configurations become dominant.  For
example, the maximum population of P$_4^{*}$ for the SPC model was observed
around $\sim 245$~K (Table~\ref{table:density}.  This is not the case for
TIP4P/2005, where the maximum is located at a 40~K larger temperature.  The
same behavior was observed comparing the temperatures at which P$_4$ and P$_3$
are equal (e.g., around 270 K for TIP4P/2005).  These observations suggested
that a temperature shift factor ($\Delta T_{s}$) exists among the models. Using
TIP4P/2005 as a reference, we found a temperature shift factor for each model
ranging from 65~K to 6~K (see Fig.~\ref{fig:deltats} and
Table~\ref{table:density}).  TIP4P/2005 was chosen as reference for its ability
to reproduce the density curve \cite{Abascal2009}. Applying this shift to the
data allowed the superposition of all models onto four master curves, one for
each structural configuration, as shown in the monochrome plot at the bottom
right of Fig.~\ref{fig:micro_pop}.  Our observation is consistent with
previously found phase diagram shifts among different water models
\cite{Sanz2004,Vega2005shift} as well as in the presence of ions
\cite{Corradini2011} but in this case we could superimpose all models onto a
master curve. Unfortunately, TIP5P had to be excluded from the superposition
because all points show an increased curvature with respect to the other
models, consistent with the increased curvature of the isobaric density at 1
atm~\cite{Vega2005}.  The structural temperature shift is larger for three-site
models (Fig.~\ref{fig:deltats}) with a spread of up to 65K for TIP3P. On the
other hand, four-site models deviate less. Both SPC-E and TIP4P are
characterized by a temperature shift with respect to TIP4P/2005 of around 20K.
In general, models providing a better estimation of the position of the density
maximum deviate less.

\begin{figure}[b]
  \includegraphics[width=80mm]{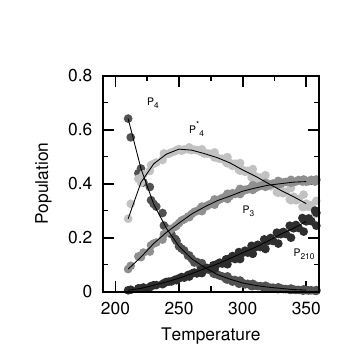}
  \caption{Overlap of the P$_i$ data when a Skinner definition \cite{Kumar2007}
  for the hydrogen bond is used.}
  \label{fig:skinner}
\end{figure}

\red{In order to check the robustness of the P$_i$ overlap with the hydrogen
  bond definition, the recent definition of Skinner and collaborators \cite
  {Kumar2007} was applied.  Fig.~\ref{fig:skinner} shows that the overlap
  between the curves is independent from the hydrogen bond definition.
  Moreover, the temperature shifts calculated in this case are very similar to
  the ones reported in Table~\ref{table:density}. It is interesting to note
  that this bond definition is very different from the one used in
Fig.~\ref{fig:micro_pop}, being based on an empirical fit of the electronic
occupancy \cite{Kumar2007}.}

\subsection{Stabilization of the fully coordinated configuration}

\begin{figure}
  \includegraphics[width=60mm]{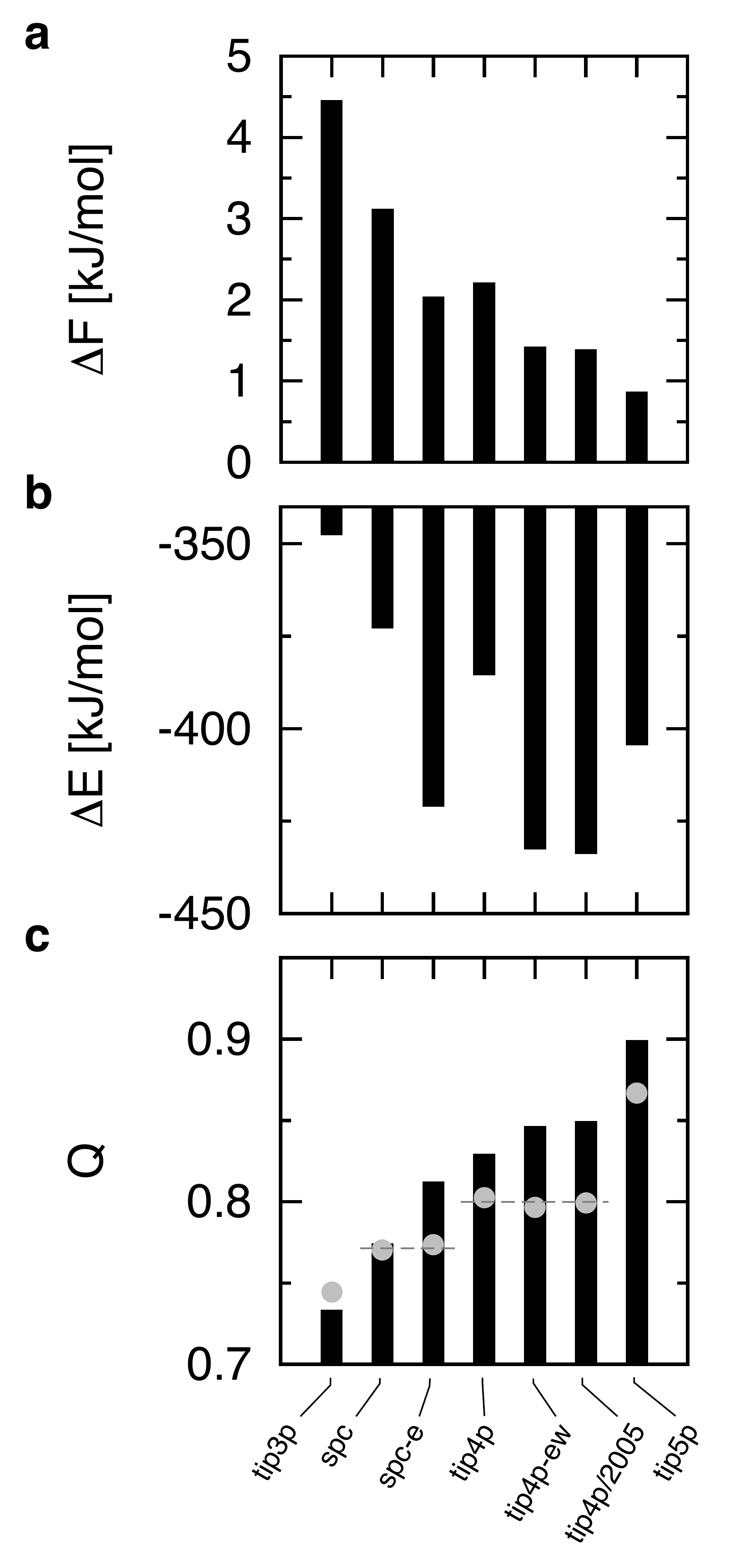}
  \caption{Comparison of water models with respect to the fully coordinated
    configuration at 230 K. ({\bf a}) The value of the free energy. ({\bf b}) Average
  enthalpy. ({\bf c}) Average value of the tetrahedral order parameter $Q$. The value
  of $Q$ for the case P$_4=0.2$ is shown as gray filled circles.}
  \label{fig:energy}
\end{figure}

At all temperatures, water models with smaller shifts provide an improved
stabilization of the fully coordinated configuration. (Alternatively, it can be
said that these models destabilize poorly hydrogen-bonded configurations).  To
make this point clearer, the free energy of the fully coordinated configuration
at 230K was calculated  (Fig.~\ref{fig:energy}a and Methods). At this
temperature P$_4$ is appreciably large for all water models. Comparison with
the temperature shifts of Fig.~\ref{fig:deltats} indicates a remarkable
correlation where even the small differences between SPC-E and TIP4P are
respected \red{(Pearson correlation coefficient of 0.99). The correlation
  decreases when considering only the enthalpy, as shown in
  Fig.~\ref{fig:energy}b (correlation of 0.94, see Methods).} Interestingly,
  enthalpy and free energy correlate very well within the same model family.
  This is particularly clear when looking at the three sites models (i.e., the
  trend for TIP3P, SPC and SPC/E), suggesting a different entropic contribution
  between three and four sites models which is systematic.

Finally, the average value of the tetrahedral order parameter $Q$
\cite{Errington2001} of the fully coordinated configuration calculated at the
same temperature is shown in Fig.~\ref{fig:energy}c. In first approximation,
the parameter correlates well with the structural shift although not as good as
the free energy (\red{correlation coefficients of 0.98}).

\begin{figure}
  \includegraphics[width=90mm]{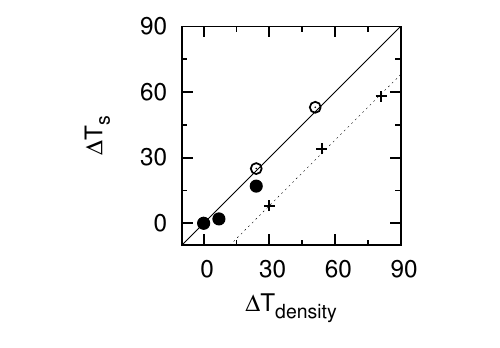}
  \caption{Comparison between the structural temperature shifts ($\Delta
   T_{s}$) and the position of the density maximum ($\Delta T_{density}$).
   Four-site models were compared to TIP4P/2005 (filled circles). Three-site
   models were compared to SPC/E (empty circles). Crosses refer to the case
  when TIP4P/2005 was used as reference for the three-site models.}
  \label{fig:tshifts}
\end{figure}

\begin{figure*}
  \includegraphics[width=120mm]{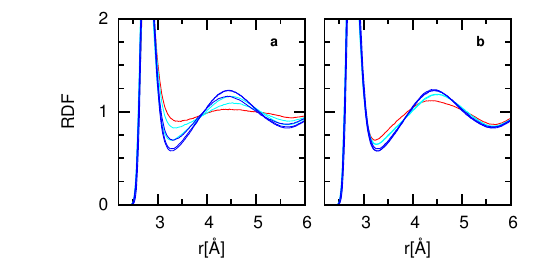} 
  \caption{Radial distribution function (RDF) ({\bf a}) at 270 K; ({\bf b}) after the
    application of the temperature shift $\Delta T_s$ (see Table I), taking
    the TIP4P/2005 model at 270~K as reference. TIP3P, SPC and TIP4P
    families are shown in red, light blue and blue, respectively.}
  \label{fig:rdf}
\end{figure*}

\subsection{Comparison with the position of the density maximum}

It is worth commenting on the relation between the structural temperature
shifts found in this work and the model-dependent temperature of maximum
density. As shown in Fig.~\ref{fig:tshifts} and Table~\ref{table:density}, the
relationship between the structural $\Delta T_{s}$ and the density $\Delta
T_{density}$ temperature shifts is linear within the three or the four-sites
models (filled and empty circles in Fig.~\ref{fig:tshifts}). However, when
comparing all models together using TIP4P/2005 as reference  a small systematic
deviation is observed (filled circles and crosses in the figure).  This is due
to the relation that exists between the populations P$_i$ and the density. It
is noted that the relative position of the P$_4^{*}$ maximum with respect to
the temperature of maximum density (dashed line in Fig.~\ref{fig:micro_pop},
see also Table~\ref{table:density}) depends on the model family.  For
four-sites models the two temperatures are identical, while for three and
five-sites models the maximum of P$_4^{*}$  is found at a higher and a lower
temperature, respectively. This behavior might be connected with the
systematic deviations between free energy and enthalpy for the different water
models (Fig.~\ref{fig:energy}a-b).

\red{To better elucidate the connection with the temperature shifts, the radial
  distribution function (RDF) was calculated. At
  270~K the RDF for the various models shows a different structural signature
  (Fig.~\ref{fig:rdf}a). As expected, only the curves for TIP4P/2005
  and TIP4P-Ew overlap, being the two models very similar. In
  Fig.~\ref{fig:rdf}b the RDF was recalculated at temperatures shifted
  according to $\Delta T_s$ using as reference TIP4P/2005 at 270~K.  The figure
  shows that all models with the same geometry perfectly overlap (e.g.  TIP4P,
  TIP4P-Ew and TIP4P/2005, blue lines), suggesting that model
  reparametrizations act as an effective shift in temperature space.
On the other hand, changes in the geometry or the number of sites affect the general shape of the
radial distribution function and consequently the density. }

  \red{Similar conclusions can be deduced when calculating the tetrahedral
    order parameter $Q$ for temperatures at which the fully coordinated
    structure has the same probability for all models (P$_4=0.2$, gray filled
    circles in Fig.~\ref{fig:energy}c). $Q$ takes the same value within a given
  family but it is influenced by the change of the molecular geometry,
indicating that the structure corresponding to fully coordinated waters depends
on the model family.}

\section{Conclusions}

\red{ In conclusion, we found that seven among the most used classical water
	models are characterized by very similar hydrogen-bond
	structure-forming capabilities up to a temperature shift.  All models
	but TIP5P perfectly overlap onto a master curve when this shift is
	applied. This behavior does not depend on the hydrogen-bond definition.
	Our findings suggest that model reparametrization acts as an effective
	shift in temperature space. On the other hand, changes in the geometry
	or the number of sites cannot be fully reconducted to temperature
	shifts alone as shown by the analysis of the density as well as the radial
	distribution function. As such, although the hydrogen bond topology is
universal when applying a certain temperature shift, this is not the case for
the structure, each model family being characterized by its own signature.}

\red{ We found that the three water models optimized to reproduce the position
  of the density maximum (i.e., TIP4P-Ew, TIP4P/2005 and TIP5P) systematically
  improve the stabilization of fully coordinated water configurations with an
  extension of at least two solvation shells.  Based on this observation, we
  speculate that the improvements of these models for biomolecular simulations
  \cite{Best2010,Nerenberg2011} are connected to the higher stabilization of
  \emph{ordered} water.  This property has important implications for the
  solvation of biomolecules, changing the balance between solute-solute and
solute-solvent interactions.}  

Development of improved force-fields strongly depends on this
balance.  Our analysis provides a microscopic and reductionist approach to
face this challenge.

\section*{Acknowledgments}

This work is supported by the Excellence Initiative of the German Federal and
State Governments.


\begin{table*}

\caption{Temperature of maximum density calculated from our simulations (TMD),
as found in the literature (TMD$^{ref}$), the structural temperature shift
($\Delta T_{s}$) and the temperature at which P$^*_4$ is maximum for the seven
water models investigated in this work. }
\begin{tabular}{p{30mm}p{30mm}p{30mm}p{30mm}p{30mm}}
  \hline
  \hline
   Water model & TMD & TMD$^{ref}$ & $\Delta T_{s}$ & $T_{max(P4^{*})}$ \\
  \hline 
   TIP3P & 199 & 182\cite{Vega2005} & 65 & 229\\
  
  SPC & 226 & 228\cite{Vega2005} & 42 & 247 \\
  
   SPC/E & 250 & 241\cite{Rick2004} & 18 & 275\\
   
   TIP4P & 256 & 248\cite{Jorgensen1983} & 20 & 268 \\
  
   TIP4P/2005 & 280 & 278\cite{Abascal2005} & 0 & 287 \\
  
   TIP4P-Ew & 273 & 274\cite{Horn2004} & 6 & 281 \\
  
   TIP5P & 282 & 285\cite{Lisal2002} &  n.a. & 269 \\
  \hline
  \hline
  \end{tabular}
  \label{table:density}
\end{table*}

\clearpage

\end{document}